\documentclass[conference]{IEEEtran}
\IEEEoverridecommandlockouts
\usepackage{cite}
\usepackage{orcidlink}
\usepackage{amsmath,amssymb,amsfonts}
\usepackage{algorithmic}
\usepackage{graphicx}
\usepackage{textcomp}
\usepackage{xcolor}
\usepackage{microtype}
\usepackage{orcidlink}
\usepackage{graphicx}
\usepackage{url} 
\usepackage{hyperref}
\usepackage{fancyhdr}

\begin{document}

\title{Approaches to Responsible Governance of GenAI in Organizations}

\IEEEspecialpapernotice{PEER-REVIEWED AND ACCEPTED IN IEEE- ISTAS 2025}

\author{
    \IEEEauthorblockN{Himanshu Joshi\textsuperscript{1} (main contact), Shabnam Hassani\textsuperscript{1}, Dhari Gandhi\textsuperscript{1}, Lucas Hartman\textsuperscript{2}}
    \IEEEauthorblockA{\textsuperscript{1}Vector Institute for Artificial Intelligence, Toronto, Canada  \\
    \textsuperscript{2}Western University, London, Canada  }
    \IEEEauthorblockA{\textsuperscript{1}himanshu.joshi@vectorinstitute.ai, \textsuperscript{2}lhartma8@uwo.ca\\}
}

\maketitle

\begin{center}
\small \textit{Project page: \url{https://res-ai.ca/}} \\
\small \textit{Github page: \url{https://vectorinstitute.github.io/responsible-gen-ai/}}\\
\small \textit{IEEE ISTAS25 conference: \url{https://attend.ieee.org/istas-2025/program/}} \\
\end{center}

\begin{abstract}
The rapid evolution and integration of Generative AI (GenAI) across industries have introduced unprecedented opportunities for innovation while also presenting complex challenges around ethics, accountability, and societal impact. This white paper draws on a combination of literature review, established governance frameworks [1-10], and insights from industry roundtable discussions with industry experts varying in professional backgrounds and organizations. Weekly discussions with these experts have contributed valuable practical insights that have enriched the paper, ensuring that its recommendations are grounded in real-world experiences and challenges. Through an analysis of existing governance models, real-world use cases, and expert perspectives, this paper identifies core principles for integrating responsible GenAI governance into diverse organizational structures.
The primary objective is to provide actionable recommendations for organizations to adopt a balanced, risk-based governance approach that allows for both innovation and oversight. Through an analysis of existing governance models, expert roundtable discussions, and real-world use cases, this paper identifies core principles for integrating responsible GenAI governance into diverse organizational structures. 
Findings emphasize the need for adaptable risk assessment tools, continuous monitoring practices, and cross-sector collaboration to establish trustworthy and responsible AI. These insights provide a structured foundation for organizations to align their AI initiatives with ethical, legal, and operational best practices.

\end{abstract}


\section{Introduction}
\subsection{Defining GenAI Governance}
GenAI governance is a structured framework of policies and practices that guides the responsible development, deployment, and oversight of AI systems. It ensures alignment with organizational values and societal expectations while managing risks such as bias, privacy concerns, and security vulnerabilities. A well-defined governance framework for GenAI fosters transparency and accountability, both essential for building trust among users and stakeholders. 
Unlike static compliance measures, responsible GenAI governance is an adaptive strategy that integrates AI applications, whether developed internally or acquired, into an organization’s long-term goals, ethical standards, and regulatory obligations. Beyond risk mitigation, effective governance enhances the efficiency, reliability, and fairness of GenAI systems throughout their lifecycle. GenAI governance in any organization should involve a layered approach that takes into account strategic, operational, and tactical considerations to foster responsible GenAI innovation\cite{huang2025ethical,batool2023responsible}. 

\subsection{Purpose and Importance of AI Governance in the Age of GenAI}
The fast-growing pace of GenAI and agentic technologies have transformed industries by enabling automation, creative content generation, and complex decision-support systems. However, these advancements introduce new risks that extend beyond traditional AI. Conventional AI primarily focuses on predictive modeling and structured data analysis, while GenAI operates in more unpredictable and dynamic contexts, often generating content that is difficult to validate or control. Growing Issues such as misinformation, intellectual property violations, data privacy,and ethical dilemmas with GenAI necessitate the requirement for stronger oversight mechanisms. 
Establishing responsible governance frameworks for GenAI~\cite{NISTAI600-1,AIRMF,AIRMF1.0,SIngaporeModelAI,AlanTuring,RAI2024,AIGovernanceBehind,AIGovernanceinPractice,ISO420012023} is essential to ensure these technologies align with organizational values, and legal and regulatory obligations while fostering innovation responsibly.

\subsection{Key Governance Challenges in GenAI}
\begin{figure}
    \centering
    \includegraphics[width=1\linewidth]{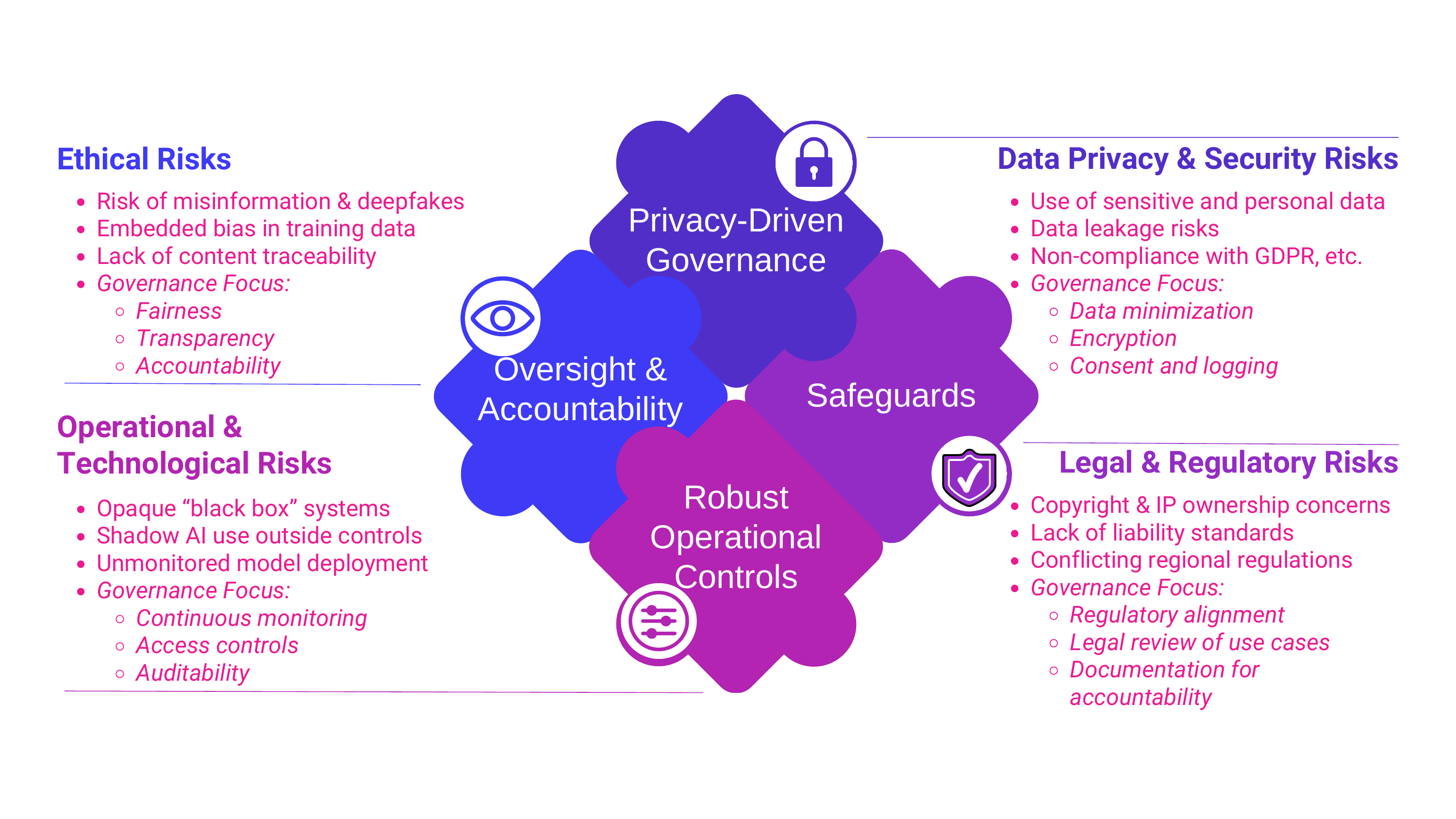}
    \vspace{-2em}
    \caption{Key Governance Challenges in GenAI}
    \label{fig:challenges}
\end{figure}

As organizations adopt GenAI, they must navigate a rapidly evolving landscape of risks and responsibilities. While the potential for automation and decision-making support is immense, these systems also pose complex governance challenges that require robust oversight frameworks. Understanding the key risks associated with GenAI is an essential start to responsible GenAI (Fig.~\ref{fig:challenges}).

\subsubsection{Ethical Risks}
One of the most significant concerns with GenAI is its ability to generate complex, high-quality content autonomously because of its possibility to contribute to misinformation, deepfakes, and bias. The difficulty in tracking and verifying AI-generated content raises serious ethical questions about its influence on public perception, decision-making, and social behavior. Addressing these risks requires governance frameworks that prioritize fairness, transparency, and accountability. Ensuring fairness involves mitigating biases in training data and model outputs, while transparency fosters clarity on how AI-generated content is created and validated. Accountability mechanisms must be embedded in order to detect, prevent, and correct potentially harmful outputs. 
\subsubsection{Operational and Technological Risks}
From a technical and operational standpoint, GenAI systems function as black boxes, i.e., they often make it difficult to interpret or audit their decision-making processes. This not only lacks transparency, but also poses challenges in critical sectors such as, healthcare, finance, and legal industries, where trust and reliability are non-negotiable. Moreover, the rise of Shadow AI (the use of unauthorized AI models outside organizational oversight) introduces significant vulnerabilities, compliance risks, and ethical concerns. Employees or teams may develop and deploy AI tools independently, bypassing any established governance controls, which can ultimately result in data leaks, regulatory violations, and the use of unreliable model outputs. To address these risks, governance frameworks must incorporate continuous monitoring mechanisms and adaptive risk management strategies that can evolve alongside advancements in GenAI. 
\subsubsection{Data Privacy and Security Risks}
GenAI’s reliance on vast amounts of training data - often collected from publicly available sources - raises serious concerns around data privacy, security, and regulatory compliance. Many GenAI models process sensitive data, including personal identifies and confidential information. Without stringent data governance protocols, these models may inadvertently expose or misuse sensitive information. Regulatory frameworks such as General Data Protection Regulation (GDPR), the EU AI Act, and sector-specific data privacy laws impose strict requirements on AI systems that process personal data. Organizations deploying GenAI must ensure that their governance strategies align with the evolving global compliance standards, emphasizing data minimization, data encryption, and robust auditing mechanisms.
\subsubsection{Legal and Regulatory Risks}
With the rapid adoption of GenAI, it has outspaced existing legal frameworks, resulting in uncertainty around intellectual property rights, liability, and compliance requirements. AI-generated content raises complex questions around copyright ownership, attribution, and fair use - particularly in creative industries where GenAI models are used to produce art, music, writing, and other types of digital content. Beyond these intellectual property concerns, organizations must also navigate sector-specific regulations that impact AI deployment. In industries such as finance, healthcare, and defense, AI systems must comply with strict laws governing data usage, bias mitigation, and accountability for decisions. Governance frameworks must proactively integrate legal expertise into GenAI risk assessments to ensure compliance with emerging AI legislation and best practices. 

\section{Related Work}

\subsection{Review of Current Frameworks}
The global landscape of AI governance has produced multiple frameworks aimed at addressing the ethical, regulatory, and operational challenges posed by AI systems, including GenAI. These frameworks provide a foundation for understanding the responsibilities and complexities involved in deploying GenAI technologies. However, while these governance models provide valuable insights, they are not universally applicable, and gaps remain in addressing GenAI's unique risks.
While several leading AI governance frameworks have emerged to address the unique challenges of GenAI, no single model fully captures the evolving risks of GenAI. Each framework offers distinct approaches to risk management, ethics, and operational transparency. The following comparative analysis examines key frameworks that have shaped the current governance landscape.

\subsubsection{Enterprise-Focused Frameworks}
\begin{itemize}
    \item NIST AI Risk Management Framework (USA): The NIST AI RMF~\cite{AIRMF,AIRMF1.0,NISTAI600-1} takes a lifecycle-based approach to AI governance, emphasizing transparency, accountability, and continuous monitoring. Its structured methodology—Govern, Map, Measure, and Manage—aligns well with enterprise risk management and provides a robust foundation for organizations navigating national AI regulatory principles.

    \item ISO AI Standards (ISO/IEC 42001): The ISO standard~\cite{ISO420012023} emphasizes a harmonized risk management framework that is applicable across industries. It provides structured compliance guidelines, particularly for organizations operating internationally or seeking certifications for AI governance.
\end{itemize}

\subsubsection{Global Perspectives on AI Governance}
\begin{itemize}
    \item Singapore Model AI Governance Framework: Singapore’s framework~\cite{SIngaporeModelAI} presents an innovation-friendly regulatory approach that balances oversight and flexibility. The model’s emphasis on fairness and explainability aligns with public trust initiatives in AI governance.

    \item Responsible AI Institute (RAI): The RAI model~\cite{RAI2024} focuses on AI certifications and compliance mechanisms, enabling organizations to demonstrate responsible AI adoption. This model is particularly beneficial for enterprises looking to establish international credibility in AI ethics and governance.
\end{itemize}
\subsubsection{Functional and Comprehensive Tools}
\begin{itemize}
    \item MIT Risk Repository: The MIT AI Risk Repository~\cite{MIT} serves as a foundational reference for AI risk categorization. By structuring risks across causal and domain taxonomies, it provides organizations with a systematic approach to risk assessment.

    \item Alan Turing Institute’s SSAFE-D Principles: The SSAFE-D framework~\cite{AlanTuring} (Safety, Sustainability, Accountability, Fairness, and Explainability) provides a process-driven approach to AI governance.
\end{itemize}

\subsubsection{Key Themes Across Frameworks}
A comparative analysis of these frameworks reveals common governance priorities.

\textbf{Continuous Monitoring and Adaptability:} Given the dynamic nature of GenAI technology necessitates adaptable governance models that can evolve over time. NIST's AI RMF, with its lifecycle-based approach, serves as a model for incorporating ongoing risk assessment and responsive governance strategies.

\textbf{Balancing Innovation with Regulation:} Striking a balance between fostering innovation and imposing effective regulation is a recurring theme. Frameworks such as Singapore's model emphasize scalable risk management to prevent over-regulation, allowing organizations the flexibility to innovate while maintaining oversight.

\textbf{Ethics, Bias, and Accountability:} The need for ethical integrity and accountability in AI systems is underscored by principles from The Alan Turing Institute and tools from RAI. These frameworks emphasize transparency and fairness, addressing biases within the AI lifecycle.

\textbf{Risk Assessment and Governance Tools:} Risk management emerges as fundamental across frameworks. Risk assessment templates, monitoring tools, and bias detection algorithms support proactive governance, helping organizations maintain secure, fair, and accountable GenAI systems.

\subsubsection{Gaps Identified in Existing Frameworks}
Current frameworks provide a strong foundation, but several gaps and areas for improvement remain.

\textbf{Granularity of Risk Assessment:
Assessment:} Most frameworks provide high-level risk categorizations but lack specific guidance on managing GenAI’s operational, ethical, and legal risks. Organizations need tailored risk management strategies that address domain-specific AI governance gaps.

\textbf{Cross-Sector Adaptability:} Current frameworks often struggle to accommodate unique sector needs. For instance, a healthcare provider must navigate stringent privacy concerns under regulations like HIPAA while addressing bias in medical diagnostics. Financial institutions might focus more heavily on cybersecurity challenges to protect sensitive financial data.

\textbf{Vendor and Third-Party AI Risks:} Managing risks associated with third-party AI tools remains an unresolved issue. More robust, practical tools are needed to evaluate and manage vendor-related risks comprehensively, including transparent documentation of third-party models and data practices.

\textbf{Lack of Clear Accountability Structures:} Many frameworks lack clear accountability for governance activities, causing confusion over stakeholder roles in implementation, monitoring, and evaluation. This clarity is crucial for organizations new to AI governance, as it prevents overlap and gaps in accountability.

\subsection{Current Governance Landscape and its Needs}
The AI governance landscape is evolving to address the unique risks and challenges of GenAI. While existing governance frameworks provide foundational guidance, the practical implementation still varies across industries, regulatory environments, and organizational scales. The complexity of GenAI risks reinforce the need for adaptable governance strategies that align with sector-specific priorities.

\subsubsection{Risk-Based Approaches to AI Governance}
A risk-based approach is central to AI governance, as highlighted by frameworks such as the NIST AI Risk Management Framework (RMF) and the EU AI Act~\cite{EUAct}. These models emphasize risk classification and tiered mitigation strategies to align with the severity and impact of AI risks.

\textbf{Sector-Specific Risk Considerations:}
\begin{itemize}
    \item Financial Services prioritize cybersecurity threats, fraud prevention, and compliance with anti-money laundering (AML) regulations.

    \item Healthcare organizations focus on bias mitigation, patient data privacy, and regulatory compliance under laws like HIPAA.

    \item Autonomous systems and defense sectors require strict safety, transparency, and accountability measures to mitigate unintended consequences in high-risk deployments.
\end{itemize}
Given these sectoral differences, organizations require flexible governance models that can adapt risk assessments and compliance strategies based on industry-specific priorities.

\subsubsection{Operationalizing GenAI Governance}
Embedding GenAI governance practices within everyday workflows is key to ensuring that governance principles are actionable. This involves:
\begin{itemize}
    \item Embedding GenAI governance within existing workflows to ensure compliance is seamless rather than a separate, burdensome process.
    \item Developing decision-support tools, such as risk matrices and GenAI lifecycle governance checkpoints, to aid AI risk evaluation at every stage—from development and deployment to post-deployment monitoring.
    \item Ensuring cross-functional collaboration between GenAI developers, compliance teams, and business leaders to operationalize governance policies without slowing innovation.
\end{itemize}
Many organizations struggle with the practical implementation of GenAI governance because existing frameworks lack industry-specific guidance. Creating adaptable, context-aware governance policies is essential for ensuring accountability without impeding technological advancements.

\subsubsection{Global Collaboration and Standardization}
The globalization of GenAI necessitates a harmonized approach to AI governance. While standards such as ISO/IEC 42001 provide a foundation for international AI governance alignment, practical adoption across jurisdictions remains a challenge due to:
\begin{itemize}
    \item Regulatory fragmentation, where different countries and industries impose conflicting compliance requirements.
    \item Variations in enforcement mechanisms, leading to inconsistent implementation across organizations operating in multiple regions.
    \item The need for clearer interoperability standards, enabling GenAI governance models to be scalable and transferable across global markets.
\end{itemize}
Achieving global GenAI governance cohesion will require cross-border collaboration between governments, industry leaders, and regulatory bodies to develop scalable, universally recognized GenAI governance protocols.

\subsubsection{Sector-Specific Adaptability in Governance Frameworks}
Industry experts highlight significant challenges in implementing governance frameworks across different sectors. Key issues include integrating GenAI into existing systems and adapting frameworks to meet the unique demands of different sectors. Recognizing that GenAI risks are often sector-specific—such as emphasizing cybersecurity in finance versus prioritizing bias mitigation in healthcare—underscores the need for a tailored, flexible governance approach that can address distinct operational and ethical concerns across industries.
\begin{itemize}
    \item Healthcare and Life Sciences:
    
        (1) Risk: Algorithmic bias in medical diagnostics could lead to discriminatory patient outcomes.
        
        (2) Governance Need: Strict bias mitigation protocols and continuous monitoring of AI-assisted decision-making.

    \item Financial Services:
    
        (1) Risk: AI-driven fraud detection systems must balance accuracy with fairness, avoiding unintended discrimination in risk assessments.
        
        (2) Governance Need: Robust model validation, bias audits, and explainability to align with financial regulations and consumer protection laws.

    \item Public Sector and Legal Compliance:
    
        (1) Risk: GenAI's use in automated decision-making and content generation could lead to misinformation or procedural injustices.
        
        (2) Governance Need: Clear GenAI accountability frameworks and mechanisms for human oversight in high-stakes AI decisions.
        
\end{itemize}
By acknowledging sector-specific governance needs, organizations can move beyond one-size-fits-all approaches and tailor GenAI governance strategies to address operational, ethical, and regulatory considerations unique to each domain.
    
\section{Identified Concerns and Risks}
As GenAI adoption expands across sectors, organizations face a spectrum of risks that demand structured governance approaches to mitigate ethical, legal, and operational challenges. Drawing from industry analyses, working group insights, and expert discussions, this section examines core risk categories, providing contextual understanding of their implications and the necessity for proactive responsible governance measures.

\subsection{Data Privacy and Integrity}
Data privacy emerges as a significant concern as GenAI models rely on vast datasets. Key challenges include:
\begin{itemize}
    \item Privacy Violations: GenAI models may inadvertently generate outputs containing private or identifiable information.

    \item Balancing Data Minimization and Model Performance: There is an ongoing challenge between data minimization principles and maintaining model accuracy, especially in high-stakes applications. While privacy laws advocate for data minimization, model performance often relies on large, diverse datasets, creating a fundamental tradeoff between privacy protection and system accuracy.

    \item Regulatory Compliance: Global privacy laws like GDPR, CCPA, and Quebec Law 25 impose strict requirements on data handling, including purpose limitation, data subject rights, and transparency mandates. However, large, unstructured datasets used in GenAI pose unique challenges for traceability and compliance monitoring.
\end{itemize}
Organizations must establish clear data governance policies that align GenAI data practices with legal and ethical standards. This includes mechanisms for de-identification, secure storage, and auditability to safeguard privacy while maintaining model reliability.

\subsection{Bias and Discrimination}
One of the most discussed concerns surrounding GenAI is its potential to perpetuate and even amplify societal biases, which can lead to discriminatory outcomes~\cite{BAIR2025}. Key challenges include:
\begin{itemize}
    \item Bias in Training Data: GenAI models learn from historical data, which may contain inherent biases. A GenAI based diagnostic tool trained on limited demographic data might generate less accurate recommendations for underrepresented patient groups, affecting treatment quality. If unchecked, these biases can influence model predictions and reinforce stereotypes.

    \item Impact on Vulnerable Populations: Biased GenAI systems can disproportionately harm historically marginalized or underrepresented groups, leading to unequal access, reduced service quality, or adverse outcomes in critical domains such as healthcare, education, finance, and criminal justice. Its consequences extend beyond just affecting the accuracy.

    \item Bias Detection and Mitigation: Organizations struggle to detect and mitigate bias, particularly when biases are hidden in proxies or encoded in complex patterns within large-scale models. Efforts to correct bias through synthetic data or model fine-tuning can also introduce new unintended biases.
\end{itemize}
Effective bias mitigation requires ongoing auditing and monitoring of GenAI models to detect and mitigate bias. Implementing Responsible AI tools, such as those developed by the Responsible AI (RAI) Institute, can help automate bias detection. However, bias and discrimination are not only ethical challenges but legal issues, with protections enshrined in frameworks like The Canadian Human Rights Act, the Ontario Human Rights Code, and the Canadian Charter of Rights and Freedoms. While technical solutions are important, effective bias mitigation must also involve diverse oversight teams and continuous alignment with both ethical principles and legal standards.

\subsection{Operational Challenges}
Integrating GenAI into established business operations introduces logistical and resource challenges that are amplified by its unique characteristics:

\begin{itemize}
    \item Continuous Model Maintenance and Drift Prevention: GenAI models require ongoing updates to prevent "drift" and maintain accuracy due to their creative nature of outputs, particularly critical in high-stakes fields. 

    \item Transparency and Explainability: GenAI models often operate as "black boxes" in applications affecting vulnerable populations. Discriminatory outcomes can influence access to services. For instance, discriminatory outcomes in healthcare could lead to disparities in treatment recommendations or diagnostic boxes," with the ability to generate novel and highly contextual outputs complicating regulatory compliance and decision transparency.

    \item Infrastructure, Resource, and Skill Demands: Deploying GenAI requires substantial computational resources and specialized expertise that many organizations may lack due to reliance on large-scale models and extensive training data.
\end{itemize}
Organizations should adopt a structured approach to operationalizing GenAI. This includes establishing dedicated teams for model monitoring and maintenance, investing in transparency tools, and providing ongoing training. Testing models in controlled environments (sandboxing) before deployment allows potential issues like biases or vulnerabilities to be identified and resolved before deployment.

\subsection{GenAI System Evaluation}
Once key risks are identified, organizations must assess GenAI systems to determine their risk levels and prioritize resources accordingly. Key challenges include:
\begin{itemize}
    \item Selection Criteria: Clear criteria for identifying high-risk GenAI systems should account for:
    \begin{enumerate}
        \item Risk Levels: From "unacceptable risk" to "minimal risk".
        \item Potential Societal Impact: Prioritizing systems affecting critical aspects of people's lives.
        
            \textbf{High Risk:} A loan eligibility system with the potential to reinforce societal inequities through biased assessments.
            \textbf{Limited Risk:} A content recommendation system where errors would have less severe consequences.
            
        \item Regulatory Exposure: Prioritize systems operating in highly regulated sectors.
        \item Frequency of Use: Assess scale and frequency of application
    \end{enumerate}
    \item Evaluation Standards: Measurable standards should be developed that align with previously outlined risks (bias, transparency, privacy). For example, a banking AI system determining loan eligibility must undergo rigorous testing for bias detection to mitigate biases that could disproportionately impact marginalized groups.
\end{itemize}
A tiered evaluation approach - from initial screening to in-depth risk assessments—should be embedded into enterprise GenAI governance frameworks. High-risk GenAI applications, such as autonomous systems or financial decision-making tools, require ongoing validation and governance oversight.

\subsection{Vendor and Third-Party Management}
The use of third-party vendors for GenAI tools is increasingly common, but it brings risks related to control, accountability, and transparency that organizations must actively manage through comprehensive governance processes. Key challenges include:
\begin{itemize}
    \item Lack of Visibility into GenAI Supply Chains: Organizations often have limited visibility into vendors' model development processes, data sources, and potential biases.

    \item Shared Liability Concerns: Complex liability allocation when GenAI systems produce harmful outputs.

    \item Compliance with Organizational Standards: Vendor models may not fully align with organizational GenAI governance policies or regulatory obligations. Additionally, Shadow AI risks emerge through unauthorized GenAI tools outside of governance structures, creating potential security vulnerabilities and compliance issues.
\end{itemize}
To mitigate vendor-related risks, it is essential for governance frameworks to implement:
(A) Due diligence processes, including risk audits and vendor impact assessments. (B) Contractual accountability provisions, specifying compliance obligations and dispute resolution mechanisms. (C) Ongoing vendor monitoring, ensuring alignment with GenAI governance policies.

\subsection{Governance and Compliance}
As regulatory landscapes evolve, organizations face challenges in maintaining compliance with international standards~\cite{BAIR2025}:
\begin{itemize}
    \item Rapid Regulatory Changes: Organizations must continuously monitor and adapt to evolving frameworks like the EU AI Act.

    \item Compliance Across Jurisdictions: Multinational organizations must navigate varying requirements across regions.Multinational organizations must navigate varying requirements across regions.

    \item Audit and Documentation: Resource-intensive requirements for thorough documentation and audit trails.
\end{itemize}
Addressing these challenges requires governance frameworks that incorporate vendor accountability measures, ensuring alignment between external GenAI systems and organizational risk policies.

\subsection{Trust and Safety}
Building public trust in GenAI requires addressing key challenges:
\begin{itemize}
    \item Misinformation and Deepfakes: Risk of synthetic content affecting credibility in journalism, education, and political discourse.

    \item Misuse Prevention: Proactive detection and prevention of malicious uses like fraud and cyber-attacks.

    \item Ethical Responsibility: Ensuring GenAI deployment considers broader social impacts.
\end{itemize}
Addressing these challenges requires governance frameworks that incorporate mechanisms to mitigate risks associated with misinformation, misuse, and ethical responsibility. Ensuring transparency, accountability, and oversight in GenAI deployment can help organizations navigate these concerns effectively.

\section{Solutions to Address Concerns}
As GenAI continues to be integrated into organizations, its unique risks and challenges demand a structured governance approach. Organizations must transition from theoretical AI governance principles to more practical and actionable strategies that ensure ethical, legal, and operational compliance.

\textbf{Building a Governance Guide:}

To effectively deploy and scale GenAI systems, organizations must adopt a multi-layered governance model that balances: risk mitigation strategies by proactively identifying and managing GenAI-related risks, operational governance through establishing policies, oversight mechanisms, and decision-making structures, and strategic scalability by ensuring governance adapts across different organizational sizes and AI maturity levels.

The governance model must also account for the diverse range of GenAI users, which may include engineers, data scientists, product managers, and non-technical end users. By ensuring inclusivity, the GenAI governance frameworks become more practical, scalable, and most importantly accessible across various roles within an organization. 

\textbf{Framework Development: A Multi-Level Approach to GenAI Governance:}

A successful responsible AI governance framework must be embedded at all levels of an organization. Clearly defined roles at each level ensures accountability, reduces risks, and aligns GenAI practices with organizational goals.

\subsection{Levels of Execution}
The success of GenAI governance hinges on its ability to be integrated at all levels within an organization. Effective governance frameworks should address GenAI concerns from a strategic level down to tactical implementation. The governance structure must incorporate roles at each level to ensure accountability and alignment across the organization.
\begin{itemize}
    \item Strategic Level: The Board of Directors, C-level executives, and senior management are responsible for establishing high-level GenAI governance policies, regulatory compliance mandates, and ethical guidelines. While they set the overarching strategy, they typically rely on GenAI governance committees, advisory councils, and external consultants for specialized expertise.

    \item Tactical Level: Business heads, VPs, and control functions are responsible for translating strategic policies into actionable governance measures. This includes overseeing GenAI implementation, ensuring regulatory adherence, and integrating GenAI risk management practices into daily operations.

    \item Operational Level: Functional managers, data scientists, and developers are responsible for executing GenAI governance practices through model development, deployment, and ongoing monitoring. They follow governance policies set at the strategic level and work closely with operational leadership to ensure compliance and risk mitigation.
\end{itemize}
Clearly defined roles at each level ensure accountability and promote a cohesive governance structure across all functional areas, enabling GenAI governance that is both comprehensive and actionable.

\subsection{Key Stakeholders}
Responsible GenAI governance is not confined to a single team; it requires cross-functional collaborative across multiple stakeholders involving:

\begin{itemize}
    \item GenAI Builders: Developers and engineers who create and maintain GenAI systems, responsible for incorporating governance principles into technical workflows.

    \item Risk and Compliance Teams: Oversee GenAI operations to maintain regulatory and ethical compliance.

    \item Business and Product Leaders: Need GenAI literacy to make informed decisions about GenAI deployment. 

    \item AI Users: End-users who interact with GenAI systems, needing guidance on appropriate and responsible use.

    \item Legal and IT Security Teams: Provide oversight on privacy risks, intellectual property, and cybersecurity.

    \item External Stakeholders: Regulatory bodies, customers, and vendors, who can impact or be impacted by the organization’s GenAI practices.
\end{itemize}
To facilitate cohesive governance, many organizations establish cross-functional GenAI councils or committees that include representatives from compliance, audit, legal, and technical teams. This council ensures GenAI initiatives align with organizational values, regulatory standards, and ethical considerations, guiding policy development and adoption.

\subsection{Bidirectional Approach: Top-Down and Bottom-Up Governance} 
A well-rounded GenAI governance model should incorporate both top-down policies and bottom-up feedback, ensuring adaptability in a rapidly evolving GenAI landscape: 
\begin{itemize}
    \item  Top-Down Governance: Strategic policies and GenAI principles are set at the executive level, providing a high-level roadmap for responsible GenAI development and 
    deployment. This approach helps in aligning GenAI initiatives with broader organizational goals, regulatory compliance, and ethical standards.
    
    \item Bottom-Up Feedback: By integrating input from developers, end-users, and operational teams, governance policies can be fine-tuned to reflect real-world challenges and opportunities. This feedback loop enables organizations to stay agile and responsive to emerging risks and technological advancements.
\end{itemize}

\textit{Practical implementation strategies include:}
\begin{itemize}
    \item Sandbox Testing Environments: Use case evaluations and sandbox testing environments provide platforms for this bidirectional flow. For example, an organization can establish a sandbox for GenAI model testing, allowing developers to assess model impacts in a controlled environment while aligning with strategic objectives.

    \item Continuous Feedback Loops: Top-level governance teams should actively solicit feedback from operational staff, refining policies based on real-world insights. Regular meetings between technical teams and executive governance committees help ensure that policies stay relevant and actionable.
\end{itemize}
This bidirectional strategy facilitates alignment between high-level policies and on-the-ground realities, creating a governance model that is resilient and adaptable.

\subsection{Foundations / Pillars of Responsible GenAI}
An effective GenAI governance framework is built on foundational pillars that support ethical, secure, and effective deployment of GenAI systems. These pillars provide structural integrity, guiding GenAI development, deployment, monitoring, and continuous improvement.

\subsubsection{Core Foundational Pillars}
Certain pillars are universally essential across organizations, forming the foundation of responsible GenAI governance. These include Ethical Practices, Data Governance, and Technical Foundations–—forming the bedrock of responsible GenAI practices. They are applicable at all levels of governance, regardless of organizational maturity or specific use cases.
\begin{itemize}
    \item Ethical and Responsible GenAI Practices: At the core of the governance framework lies a strong ethical foundation. Governance in GenAI must ensure that the technology, processes, and outcomes align with ethical standards, legal obligations and organizational expectations. This foundation includes essential practices for bias mitigation, privacy, and security. Ethical GenAI practices are necessary to foster trust and compliance, ensuring that GenAI systems serve humanity in equitable and inclusive ways.

    \item Data Governance and Privacy: Effective GenAI governance relies on a robust data strategy. This pillar ensures that data is accurate, representative, and managed responsibly. Key aspects include data lineage (tracking the flow of data across systems), model lineage (tracking model evolution and dependencies), and ensuring data is free from bias. Proper data governance guarantees that data quality is maintained, which is crucial for the transparency and reliability of GenAI outputs. By implementing proper data governance, organizations can align data practices with privacy requirements and support responsible data utilization.

    \item GenAI and Data Literacy/Education: Beyond technical training, a culture of GenAI and data literacy,  alongside ethical education is essential. This pillar aims to build an understanding across the organization of what constitutes responsible GenAI and data use. Employees at all levels, from HR managers to legal advisors, must be aware of the limitations, risks, and ethical considerations associated with GenAI and data. This literacy helps prevent inappropriate usage, such as misinterpreting data, mishandling sensitive information, or uploading confidential content into GenAI tools like ChatGPT.

    \item Use Case Evaluation and Sandbox Environments: Innovation within safe boundaries is facilitated by use-case evaluations and sandbox environments. Sandbox environments enable employees and departments to test GenAI applications within controlled environments, allowing for experimentation without risking data privacy or security. This structure supports the containment of new ideas, addressing concerns around Shadow AI by providing a safe internal environment for development rather than relying on external, uncontrolled tools. To further prevent the risks of Shadow AI, organizations should implement access controls, monitoring mechanisms, and usage policies that restrict employees from using external GenAI tools for sensitive tasks, such as writing code, processing proprietary data, or generating official business content. Additionally, educational initiatives can reinforce the risks of uncontrolled GenAI tool usage and promote best practices for responsible AI experimentation within secure environments.
\end{itemize}

\subsubsection{Supporting Pillars for Operational-Level Execution}
These pillars, while essential, can be adapted to fit the unique operational needs, maturity level, and use cases of each organization. They are designed to support the core foundational elements and provide additional layers of accountability and operational flexibility.
\begin{itemize}

    \item GenAI Risk Management: This pillar focuses on proactive risk identification and mitigation strategies, utilizing external repositories (e.g., MIT’s Risk Repository) and internal data sets to address potential GenAI risks.
    
    \item Security and Infrastructure: Security in GenAI governance encompasses cybersecurity, infrastructure security, and operational security. This pillar ensures that data centers and GenAI infrastructure are secure from both physical and digital threats. The security focus should include personal data privacy, and operational security measures, ensuring the protection of both data and GenAI systems in compliance with global security standards.
    
    \item Regulatory Compliance and Auditing: Compliance with regulatory and legal requirements is essential for maintaining GenAI governance integrity. This pillar includes regular audits and reporting mechanisms to track data, model performance, and adherence to governance policies. An effective compliance structure ensures transparency and accountability, enabling organizations to align with both internal and external standards.
    
    \item Control and Reporting: To maintain oversight and ensure policy implementation aligns with organizational goals, organizations must establish clear control and reporting structures. This includes mechanisms to track data sources, model versions, and operational metrics, providing necessary oversight, especially in regulated sectors. These structures enable continuous monitoring of GenAI systems and alignment with governance standards, supporting accountability at all levels.
    
    \item Operational Efficiency and Training: As GenAI technologies evolve, so too must the skills and practices within the organization. This pillar emphasizes continuous training and change management, ensuring that personnel can adapt to new tools, technologies, and compliance requirements. Operational efficiency also includes implementing Standard Operating Procedures (SOPs) and best practices that enable teams to work effectively within governance constraints.
    
    \item Evaluation Toolkits: This pillar includes developing neutral evaluation tools to assess GenAI models and technology providers. Organizations should have frameworks to guide decisions on whether to use a specific model or platform (e.g., GPT-4 vs. Gemini) or to evaluate vendors (Microsoft, Google, AWS) based on compliance, performance, and alignment with organizational goals. Evaluation toolkits ensure consistent and objective decision-making.
    
    \item Trust and Safety: Trust and safety principles serve as fundamental values that are embedded throughout the GenAI governance framework. This pillar is responsible for ensuring that GenAI systems operate within ethical and safety boundaries, supporting a culture of GenAI trust and responsibility across the organization.
    
    \item Good Technical Practices: The governance framework should promote sound engineering and technical practices. This includes decisions on whether to use on-premise or cloud solutions and other technical choices that affect data security and model integrity. Good technical practices ensure that GenAI systems are reliable and aligned with the organization's standards.
    
    \item Continuous Monitoring and Improvement: Integrated continuous monitoring supports every phase of the GenAI lifecycle—beginning with use case identification, ideation, and design through to development, deployment, and post-deployment evaluation. This monitoring ensures that both the GenAI models and the governance practices remain aligned with emerging regulatory standards, technological advancements, and evolving organizational goals. For effective monitoring, it is crucial that humans involved in the GenAI governance process have a sufficient understanding of how GenAI systems function. This knowledge enables them to exercise discretion in identifying issues, making informed decisions, and intervening when necessary. Also, governance frameworks must empower individuals to report concerns or initiate changes themselves when GenAI systems deviate from expected outcomes or pose unintended risks. A focus on continuous improvement enables organizations to adapt proactively, maintaining effective and relevant governance structures that foster responsible and sustainable GenAI innovation. 
    
    \item Accountability: Finally, accountability is a cornerstone of responsible GenAI governance. There must be clearly defined roles and responsibilities at every level, with oversight mechanisms that hold stakeholders accountable for GenAI actions. As GenAI governance evolves, accountability structures should also adapt, possibly including new roles such as a Chief AI Officer (CAIO) who is distinct from the Chief Data Officer (CDO), Chief Technology Officer (CTO), or Chief Information Security Officer (CISO). The Chief AI Officer (CAIO) would be responsible for overseeing GenAI strategy, compliance, and risk management across the organization. This role would ensure GenAI systems align with ethical guidelines, regulatory standards, and organizational objectives. 
\end{itemize}
To ensure sustained focus on each pillar, organizations can establish specialized committees or dedicated roles. For instance, a "Data Governance Committee" could oversee all data-related aspects, while a "Risk Management Office" would be responsible for monitoring and managing GenAI-specific risks. Assigning responsibility for each pillar enables deeper expertise and consistent oversight. This structure fosters accountability and ensures that governance measures are continually enforced across the organization.

\subsection{Embedding Governance Across the AI Lifecycle}
An effective GenAI governance framework must incorporate the entire GenAI  lifecycle, from ideation through deployment to continuous monitoring. Each lifecycle phase acts as a governance checkpoint, ensuring foundational and supporting principles are consistently applied to manage risks, uphold ethical standards, and maintain accountability. This lifecycle model provides a comprehensive view of governance, reinforcing key practices at critical stages and enabling a structured approach to lifecycle management.

\begin{figure*}
    \centering
    \includegraphics[width=1\linewidth]{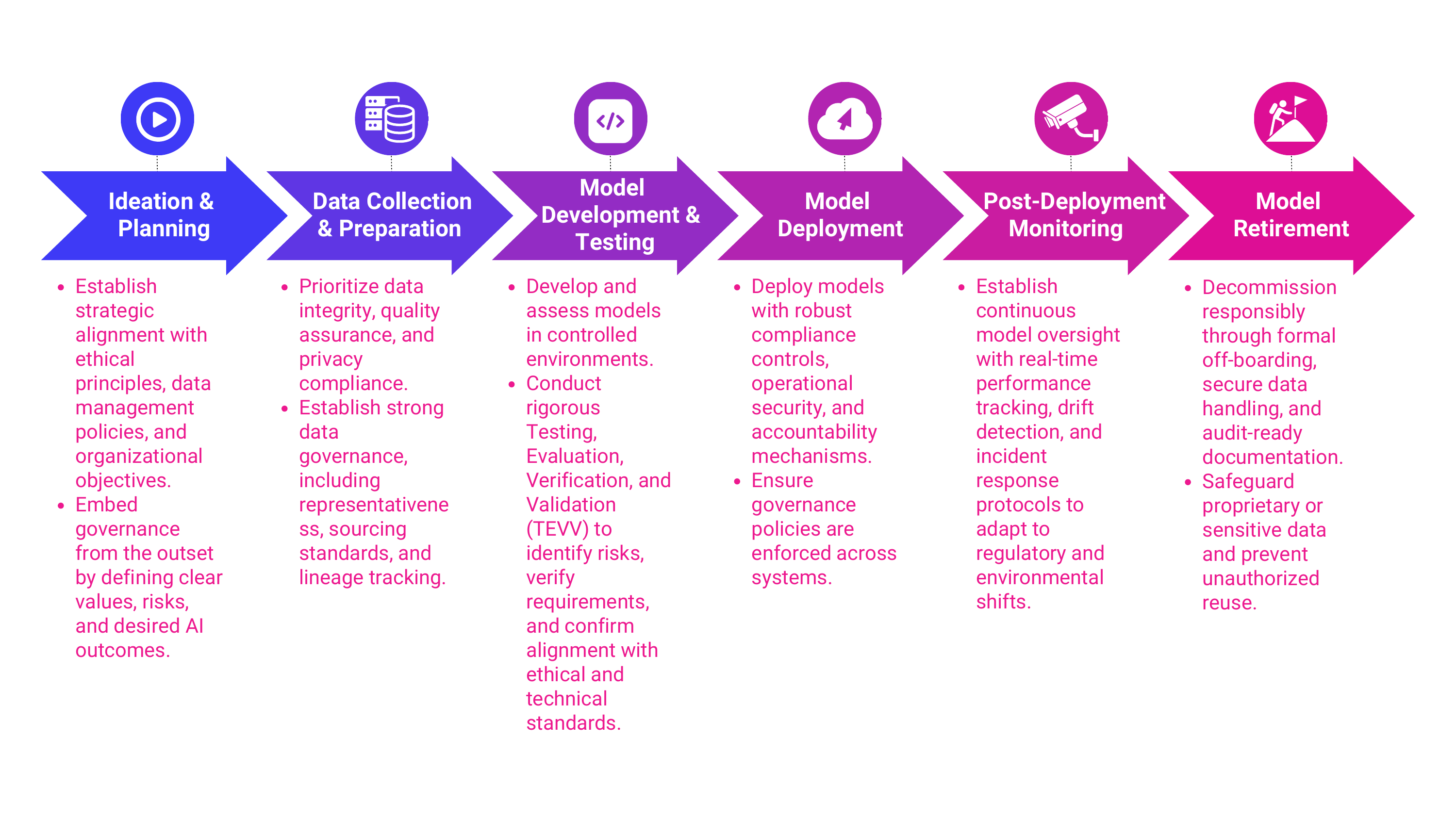}
    \vspace{-3em}
    \caption{Responsible GenAI Governance Across the Model Lifecycle}
    \label{fig:Lifecycle}
\end{figure*}

Figure~\ref{fig:Lifecycle} outlines six critical stages in the GenAI development lifecycle, each requiring distinct governance measures to ensure responsible GenAI deployment. These stages are further detailed below.

\textbf{GenAI Lifecycle Governance Stages:}
\begin{enumerate}
    \item Ideation and Planning: During ideation and planning, governance focuses on strategic alignment with ethical standards and data management principles. At this foundational phase, the organization’s core values and ethical commitments are embedded in the GenAI project’s purpose, objectives, and design. Clear guidelines on ethical GenAI practices and data governance establish a strong foundation for responsible GenAI development.
    
    \item Data Collection, Exploration, and Preparation: At the data collection and preparation stage, the governance framework prioritizes data integrity, privacy, and security, essential for responsible GenAI development. Data integrity refers to the accuracy, consistency, and reliability of data throughout its lifecycle, ensuring that information remains unaltered and trustworthy across collection, storage, processing, and analysis. Governance ensures data is representative, accurate, and responsibly sourced, particularly in regulated sectors. Robust data lineage and quality assurance practices enhance transparency and address potential biases early on, promoting equitable GenAI outcomes.
    
    \item Model Development and Testing, Evaluation, Verification, and Validation (TEVV): In the model development and testing phase, governance activities focus on risk management, ethical oversight, and rigorous assurance processes. To ensure clarity and effectiveness, this phase can be broken down into two distinct components: 

    \begin{itemize}
        \item Experimentation and Model Development: During experimentation, models are built and iteratively improved in controlled environments. These controlled settings, such as sandboxes, enable secure experimentation while also fostering innovation. This stage is critical for identifying initial design flaws, refining model objectives, as well as addressing early-stage biases/technical risks.
    
        \item Testing, Evaluation, Verification, and Validation (TEVV): Governance frameworks in this phase focus on the systematic assessment of models through structured TEVV protocols. 
        (1)Testing: Evaluating models under diverse conditions to assess performance, robustness, and fairness. 
        (2) Evaluation: Reviewing model behavior to ensure compliance with ethical and technical standards. 
        (3) Verification: Ensuring that models meet predefined requirements and specifications. 
        (4) Validation: Confirming that models align with intended outcomes and perform safely in the expected contexts. 
    \end{itemize}
    Governance activities during TEVV should also incorporate risk management protocols to identify and mitigate biases, ethical concerns, and technical vulnerabilities. These protocols align models with organizational ethical standards and ensure readiness for real-world deployment.
    
    By separating experimentation from TEVV, governance frameworks can better address the unique requirements of each stage, ensuring that both innovation and compliance are effectively managed during model development.
    
    \item Deployment: Upon deployment, the governance framework shifts to emphasize security, compliance, and accountability, ensuring models are integrated responsibly within operational systems. Compliance checks confirm adherence to governance policies, protecting organizational standards and mitigating security risks. Regular audits and reporting mechanisms enhance transparency for both internal and external stakeholders.
    
    \item Post-Deployment Monitoring and Maintenance: In the post-deployment phase, continuous monitoring and improvement become the focus. The governance framework supports a feedback loop for real-time adjustments to the GenAI system as it operates live. By embedding continuous monitoring, organizations ensure the model’s performance remains aligned with evolving regulations and technological advancements, reinforcing trust and safety. Additionally, this vigilant oversight helps detect and address concept drift—a phenomenon where the model's predictions become less accurate over time due to changes in underlying data patterns or external conditions. Proactively managing concept drift not only sustains model accuracy but also aligns with previously mentioned risk-mitigation strategies, ensuring consistent and reliable outcomes in dynamic environments.
    
    \item Model Retirement: Finally, in the retirement phase, governance focuses on responsible decommissioning. This phase emphasizes accountability and data governance, ensuring sensitive information is safeguarded and data handling complies with organizational and regulatory standards. Clear protocols govern data transfer and model offboarding, preventing unauthorized use and securing critical information.
\end{enumerate}
By mapping each lifecycle stage to specific governance pillars, GenAI governance becomes a continuous practice, adaptable as projects evolve. Embedding governance into each lifecycle phase ensures ethical, secure, and transparent practices throughout GenAI development and deployment, fostering responsible GenAI innovation that aligns with regulatory standards and organizational goals.

\subsection{Lens for Effective Adoption: Scaling Governance Across Organization Types}
GenAI governance frameworks must be flexible to accommodate the varying needs of different organization types. Large corporations and small to medium enterprises (SMEs) often have different resources, risk profiles, and operational needs, which require customized approaches to governance.

\textbf{Adopting the Framework: Large Organizations:}

For large organizations with complex structures and diverse GenAI applications, a multi-layered governance framework is essential. These organizations require detailed risk management processes, frequent compliance checks, and extensive documentation. Key considerations for large corporations include:

\begin{enumerate}
    \item Defined Governance Layers: Large corporations benefit from clearly delineated roles at strategic, operational, and tactical levels. Each layer has specific responsibilities, with distinct features: the strategic layer focuses on regulatory alignment and high-level risk assessments; the operational layer ensures implementation of governance policies across business units; and the tactical layer addresses on-the-ground deployment, model monitoring, and technical adjustments.
    
    \item Automated GenAI monitoring Systems: Comprehensive governance toolkits, including automated monitoring systems, bias detection tools, and robust compliance protocols, support large organizations in managing risks across multiple GenAI applications. These tools facilitate real-time risk assessment, enable impact measurement, and provide essential documentation for transparency and accountability.
    
    \item Regular Audits and External Oversight: Large organizations benefit from periodic audits and external oversight to validate GenAI systems, enhance public trust, and ensure regulatory compliance. Regular compliance checks help ensure alignment with both internal and external governance standards, enabling proactive risk management.
    
    \item Prioritizing High-Risk Areas: Due to their resources and complex operational needs, large organizations can afford to address multiple pillars simultaneously. Prioritizing high-risk areas—such as data governance, GenAI risk management, and control and reporting—enables an effective approach to GenAI governance, mitigating operational risks.
\end{enumerate}

\textbf{Adopting the Framework: Small and Medium Enterprises (SMEs):}

\begin{figure*}
    \centering
    \includegraphics[width=1\linewidth]{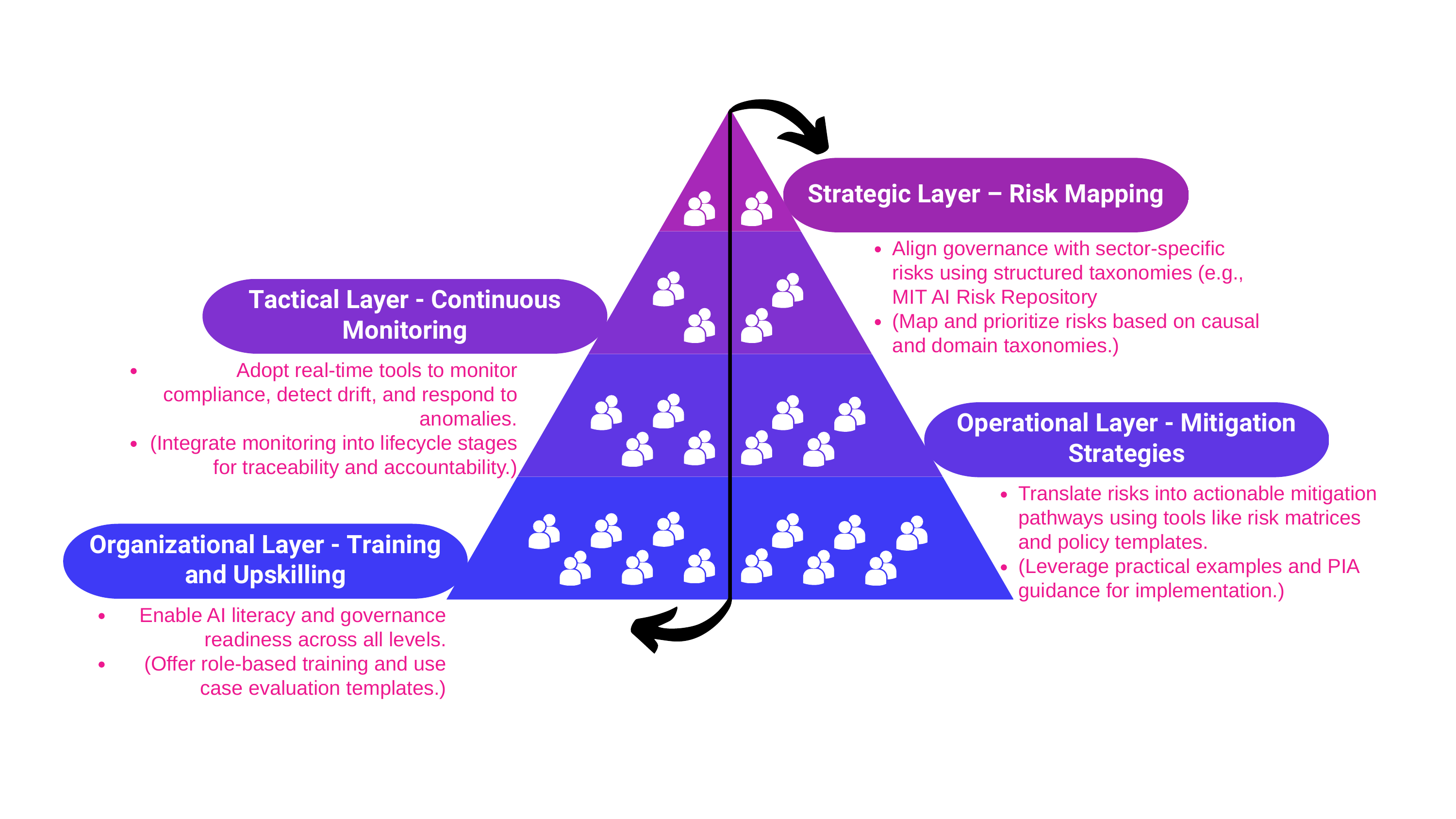}
    \vspace{-5em}
    \caption{Implementing GenAI Governance: A Bi-Directional Execution Guide}
    \label{fig:Guide}
\end{figure*}

For SMEs with limited resources, governance frameworks need to be simplified but still effective. Prioritizing essential governance elements allows these organizations to manage GenAI responsibly without overwhelming their operational capacity. As GenAI use grows within SMEs, this foundational approach can expand, aligning governance practices with larger, more sophisticated frameworks. Key considerations for SMEs include:
\begin{itemize}

    \item Streamlined Governance Layers: SMEs can begin by focusing on the strategic and operational layers of governance, emphasizing core policies and operational practices. Initially, the framework can be adapted to focus on the most critical areas, gradually building up complexity as the organization matures in GenAI usage and governance needs evolve.
    \item Focus on Core Pillars: SMEs should prioritize foundational pillars, such as data integrity, ethical GenAI practices, and basic compliance. This targeted approach allows SMEs to manage GenAI risks effectively, ensuring alignment with regulatory expectations without requiring the depth of resources that large organizations need.
    \item Practical Governance Tools: Simplified governance models tailored to SMEs include basic compliance checklists, data management templates, and ethical training programs. These tools offer SMEs a practical entry point for effective GenAI governance, ensuring that even smaller organizations maintain ethical and responsible GenAI practices from the outset.
    \item Scalability for Growth: As GenAI capabilities expand, SMEs can incrementally adopt additional pillars, such as formal risk management processes or advanced monitoring tools. This staged approach enables SMEs to scale their governance practices incrementally in line with increasing GenAI adoption, evolving towards a more comprehensive governance structure that meets their growing operational needs.
\end{itemize}
By offering scalability and customization, governance frameworks should help organizations of all sizes establish effective governance, ensuring that GenAI practices align with their resources, maturity, and operational goals. This flexible approach allows both large corporations and SMEs to integrate responsible GenAI governance as a sustainable part of their strategy.

\section{Implementation Plan: Toward Actionable GenAI Governance} \label{Implementation}
Creating an effective and actionable GenAI governance guide requires not only a solid conceptual foundation but also a structured approach that translates high-level governance concepts into operational workflows. This section outlines the practical steps and considerations for implementing the GenAI governance guide, focusing on the tools, processes, and scalability necessary to make GenAI governance a sustainable part of organizational strategy. 

A key resource in this process is the Principles in Action (PIA) framework~\cite{VectorPIA}, developed by the Vector Institute. The PIA is an interactive playbook that translates high-level AI governance principles into actionable, real-world strategies. Available at [7], the PIA serves as both a reference and a toolkit, offering actionable examples, use-case templates, and best practices to support organizations in implementing responsible AI governance. By integrating the PIA’s practical insights into the responsible GenAI governance guide, the gap between theoretical principles and operational realities can be bridged.

The GenAI governance guide focuses on practical approaches of responsible development and deployment of GenAI solutions based on ethical compliance, accountability, risk mitigation, and continuous improvement.~\ref{fig:Guide}
The guide is divided into four main execution levels (Fig.~\ref{fig:Guide}): 1)~strategic, 2)~tactical, 3)~operational, and 4)~organizational—each with distinct roles, responsibilities, and tools.

At its core, this guide focuses on the following objectives:
\begin{itemize}
    \item Ensuring Ethical and Regulatory Compliance: Aligning GenAI practices with legal obligations, ethical considerations, and global standards.
    \item Facilitating Accountability and Transparency: Enabling clear accountability structures and transparent decision-making.
    \item Mitigating Risks: Providing risk management practices across the GenAI lifecycle to prevent biases, data breaches, and unintended consequences.
    \item Promoting Continuous Improvement: Supporting adaptive governance that evolves as GenAI technologies and regulatory landscapes change.
\end{itemize}
To facilitate effective implementation, the framework will incorporate modular components, including a risk repository, evaluation toolkits, and feedback mechanisms. Each component is designed to function independently or in coordination with others, allowing organizations to scale their governance practices according to their maturity and GenAI adoption levels. The PIA document serves as a practical reference throughout, enhancing each step with actionable guidelines, decision-making frameworks, and principles grounded in responsible GenAI practices. 

\subsection{Step 1: Mapping Existing Risk Frameworks}
The first step focuses on aligning the governance framework with established risk standards and frameworks, such as the MIT AI Risk Repository and the NIST AI Risk Management Framework. While these risk frameworks provide a solid foundation for identifying and categorizing risks through structured taxonomies, the AI risk mapping builds on these insights to operationalize them. By helping organizations prioritize, analyze, and address risks specific to their industry, operational context, and GenAI lifecycle stages, this mapping creates a bridge between high-level frameworks and actionable strategies. The result is a structured foundation that organizations can adapt to their unique requirements, ensuring a clear and contextualized approach to risk identification and prioritization.

\subsubsection{GenAI Risk Mapping Tool}
We have developed a GenAI Risk Mapping tool, an extension of the foundational principles and insights provided by the MIT AI Risk Repository. This tool not only builds upon the repository's comprehensive catalog of 1600+ AI risks but also tailors and expands its applicability for diverse organizational contexts and governance requirements. While the MIT repository provides a static reference for AI risks, the GenAI Risk Mapping tool transforms these insights into actionable strategies, equipping organizations with practical resources for real-world implementation.

By leveraging the GenAI Risk Mapping tool, organizations can classify and analyze risks using both causal and domain taxonomies, enabling precise identification and prioritization. This tool allows organizations to filter risks based on their specific industry, operational level, and risk profile. Using the repository as a foundation, the framework categorizes risks under critical taxonomies, such as discrimination, data security, ethical concerns, and model failure, while introducing enhancements that address emerging and sector-specific challenges.
\begin{itemize}
    \item Causal Taxonomy: The tool organizes risks based on their origin (e.g., human error, technical faults, or malicious intent), intent (e.g., willful misuse versus unintentional consequences), and timing (e.g., pre-deployment, deployment, or post-deployment). This structure enables organizations to anticipate and address risks at every stage of the GenAI lifecycle:
    
        (1) Post-Deployment Risks: Includes issues such as model drift or adversarial attacks that emerge after systems are operational.
        (2) Pre-Deployment Risks: Focuses on challenges like data integrity and training biases that could affect downstream GenAI outputs.
        
    \item Domain Taxonomy: Organizes risks into broader categories—such as privacy, ethical GenAI, governance failures, and operational challenges—while capturing sector-specific nuances like HIPAA compliance in healthcare or operational integrity in finance. For instance:
    Competitive Dynamics: Captures risks arising from "GenAI races," where rapid deployment may compromise safety or ethical standards.
    Supplier Management: Highlights the challenges of managing third-party GenAI tools, such as inadequate transparency or oversight.
    \end{itemize}
These taxonomies empower organizations to align their governance strategies with the most pressing concerns in their respective sectors. For example:
\begin{enumerate}
    \item Telecommunications Companies: May prioritize GenAI risks related to misinformation detection, ensuring fair and unbiased content moderation on digital platforms.
    \item Energy and Utilities: Must address GenAI-driven forecasting risks, ensuring that automated grid management does not disproportionately impact specific regions or demographics.
\end{enumerate}
The PIA document further complements the GenAI Risk Mapping tool, offering real-world scenarios and use cases to demonstrate how organizations can apply these frameworks effectively. For example:
\begin{enumerate}
    \item Scenario-Based Risk Mapping: Organizations can examine practical examples, such as deploying a chatbot in a regulated industry, to identify relevant risks and appropriate mitigation strategies.
    \item Lifecycle Integration: The framework ensures that risk mapping is embedded throughout the GenAI lifecycle, from ideation and development to deployment and eventual decommissioning.
\end{enumerate}  
This mapping tool not only establishes a foundational approach but also introduces a scalable and adaptable framework, ensuring that organizations remain responsive to emerging risks and regulatory shifts. By addressing limitations in traditional frameworks, the GenAI Risk Mapping tool emphasizes adaptability and continuous improvement:
\begin{enumerate}
    \item It identifies new risks, such as GenAI arms races and the potential compromises in safety or ethics due to competitive pressures.
    \item It underscores the importance of managing risks related to third-party vendors and black-box GenAI systems, ensuring transparency and accountability across the supply chain.
\end{enumerate}
By leveraging insights from the repository, organizations can operationalize risk mapping through a variety of tailored tools and processes.

\subsection{Step 2: Incorporating Mitigation Strategies}
With risks effectively mapped and categorized in Step 1 using the GenAI Risk Mapping tool through insights from the PIA document and MIT AI Risk Repository, the next logical step is to translate this understanding into structured mitigation strategies. This ensures that identified risks are not only documented but actively managed across the GenAI lifecycle. By combining the actionable capabilities of the risk tool with practical examples from PIA, organizations can develop proactive, real-time solutions to address high-priority risks effectively. Expert feedback highlighted the importance of making this step actionable, allowing organizations to efficiently identify and address high-priority risks.

\subsubsection{From Mapping to Action: Operationalizing Risk Insights} 

The GenAI Risk Mapping tool serves as the critical link, transforming theoretical risk identification into targeted strategies for effective management. By aligning identified risks with mitigation pathways, this tool operationalizes governance frameworks and ensures that risk management is seamlessly integrated into organizational workflows. For example, risks identified during pre-deployment, such as biases in training data, can be directly addressed through corrective actions like dataset re-balancing or algorithmic adjustments. Similarly, post-deployment risks, such as model drift or adversarial vulnerabilities, are tied to continuous monitoring strategies and contingency plans. This integration moves risk management beyond documentation, embedding it into the operational reality of GenAI systems.
\begin{itemize}
    \item Continuous Monitoring Tools: Organizations should adopt automated tools for ongoing monitoring of model behavior, data integrity, and policy compliance. These tools facilitate early detection of deviations, ensuring rapid response to any governance issues that may arise post-deployment.The GenAI Risk Mapping tool enhances this process by providing insights that allow organizations to deploy automated tools to:
    
        (1) Track model behavior and performance: Detect anomalies such as declines in accuracy, fairness, or other critical metrics.
        (2) Ensure compliance: Monitor adherence to regulatory requirements and organizational policies, with the flexibility to adjust as laws or standards evolve.
        (3) Generate real-time alerts: Respond quickly to emerging threats, including adversarial attacks, data breaches, or operational failures.
        
    \item Risk Matrices: To operationalize risk prioritization, the guide includes risk matrices that categorize risks by their likelihood and impact. High-impact, high-likelihood risks are prioritized for immediate action, allowing for efficient allocation of resources. This structured visualization allows organizations to quickly compare risks across different categories and allocate resources efficiently.
    
        (1) Focused mitigation efforts: By identifying high-risk areas in the matrix, organizations can immediately target concerns such as algorithmic bias in decision-making systems or cybersecurity vulnerabilities.
        (2) Efficient resource allocation: The structured format of the matrix enables decision-makers to strategically distribute efforts, ensuring the most severe threats are addressed first while lower-risk issues are monitored accordingly.
\end{itemize}     
The PIA document emphasizes practical risk mitigation strategies, offering templates for creating risk matrices and continuous monitoring dashboards. By integrating these templates, the GenAI governance framework becomes more actionable, enabling organizations to adopt industry-best practices for monitoring and responding to potential governance issues. For instance:
\begin{itemize}
    \item A financial institution deploying credit scoring algorithms can leverage PIA resources to validate transparency and fairness, minimizing the risk of biased outcomes.
    \item A healthcare provider can apply PIA-driven strategies to ensure compliance with privacy laws while maintaining diagnostic accuracy.
\end{itemize}    
By establishing structured mitigation pathways and integrating real-time monitoring, this step helps organizations not only track but also manage risks dynamically, ensuring that governance efforts evolve in line with operational needs and emerging threats. The adaptability of the GenAI Risk Mapping tool ensures that pathways remain relevant, dynamically evolving to address emerging risks from technological advancements or regulatory changes. Through combining the GenAI Risk Mapping tool’s robust risk classification system with the actionable insights from the PIA document, organizations can create a effective framework for mitigating both current and future risks. This dual approach ensures sustainable, responsible GenAI governance that is not only effective today but also adaptable to the challenges of tomorrow.

\subsection{Step 3: Training and Up-skilling}
The final step in the initial implementation process is to develop and integrate continuous training and up-skilling programs to build GenAI literacy and ensure organizational readiness. Emphasis is placed on preparing all levels of personnel—from C-level executives to operational teams—with the knowledge and skills necessary to support responsible GenAI use and governance:
\begin{itemize}
    \item Use Case Evaluation: This modular toolkit includes standardized templates for evaluating specific GenAI use cases, focusing on assessing potential risks, regulatory requirements, and ethical considerations. The toolkit’s templates are adaptable, allowing organizations to modify them based on the complexity and risk level of each use case. By supporting a consistent evaluation approach, the toolkit aids organizations in aligning new GenAI applications with established governance practices.
    
    \item Role-Based Training Modules: Tailored training programs for executives, operational managers, and technical staff ensure that each level of the organization is equipped with relevant governance knowledge. Topics include GenAI ethics, privacy laws, compliance standards, and technical risk management.
    
    \item Building a Culture of GenAI Literacy and Adaptability: To sustain governance efforts, organizations must embed GenAI literacy into their core operations. Continuous training ensures that teams remain informed about evolving risks, technological advancements, and regulatory updates. It also ensures that employees at all levels understand their role in maintaining ethical and responsible GenAI practices and cross-functional collaboration is enhanced as personnel share a common understanding of governance principles and risk management strategies.
\end{itemize}   
The combination of insights from the GenAI Risk Mapping tool and PIA document enables organizations to create adaptive training programs that evolve alongside governance needs. For example:
\begin{itemize}
    \item  A retail organization deploying GenAI-driven recommendation systems might use the training framework to educate teams on consumer privacy risks and bias detection.
    \item  A manufacturing firm automating supply chains can train staff on monitoring GenAI for operational inefficiencies or security vulnerabilities.
\end{itemize}    
The GenAI Risk Mapping tool adds depth to training programs by bridging theoretical risk frameworks with practical applications. It provides staff with the ability to:
\begin{itemize}
    \item Understand risks comprehensively: Training modules incorporate insights from the tool, helping personnel identify and assess risks specific to their roles.
    \item Develop targeted mitigation strategies: Teams learn to link risks with appropriate mitigation pathways, ensuring alignment with organizational governance standards.
    \item Apply risk-driven decision-making: Personnel are trained to use the tool to prioritize and act on risks based on impact and likelihood, improving strategic responses across all levels.
\end{itemize}      
The PIA document provides guidance on fostering a culture of AI literacy and ethical awareness, with an emphasis on real-world applications and decision-making practices. By continuously refining training methodologies and incorporating feedback, organizations ensure their teams are not only prepared to manage current challenges but also equipped to adapt to future risks. This dynamic approach builds a foundation for sustainable and responsible GenAI governance that is rooted in knowledge, collaboration, and proactive risk management.

\section{Conclusion: Making GenAI Governance a Continuous, Scalable Process}
Implementing responsible GenAI governance requires an approach that is structured yet flexible, scalable yet practical. By following this three-step implementation plan, organizations can ensure that GenAI governance is not just a compliance requirement but an integrated part of GenAI strategy and operations.
As GenAI adoption accelerates, organizations must continuously refine governance practices, integrate real-time risk monitoring, and align GenAI strategies with evolving regulatory landscapes. By embedding governance into the AI lifecycle, corporate strategy, and organizational culture, businesses can harness the transformative power of GenAI while ensuring ethical, transparent, and responsible GenAI deployment.

\begin{figure}[h]
    \centering
    \includegraphics[width=0.75\linewidth]{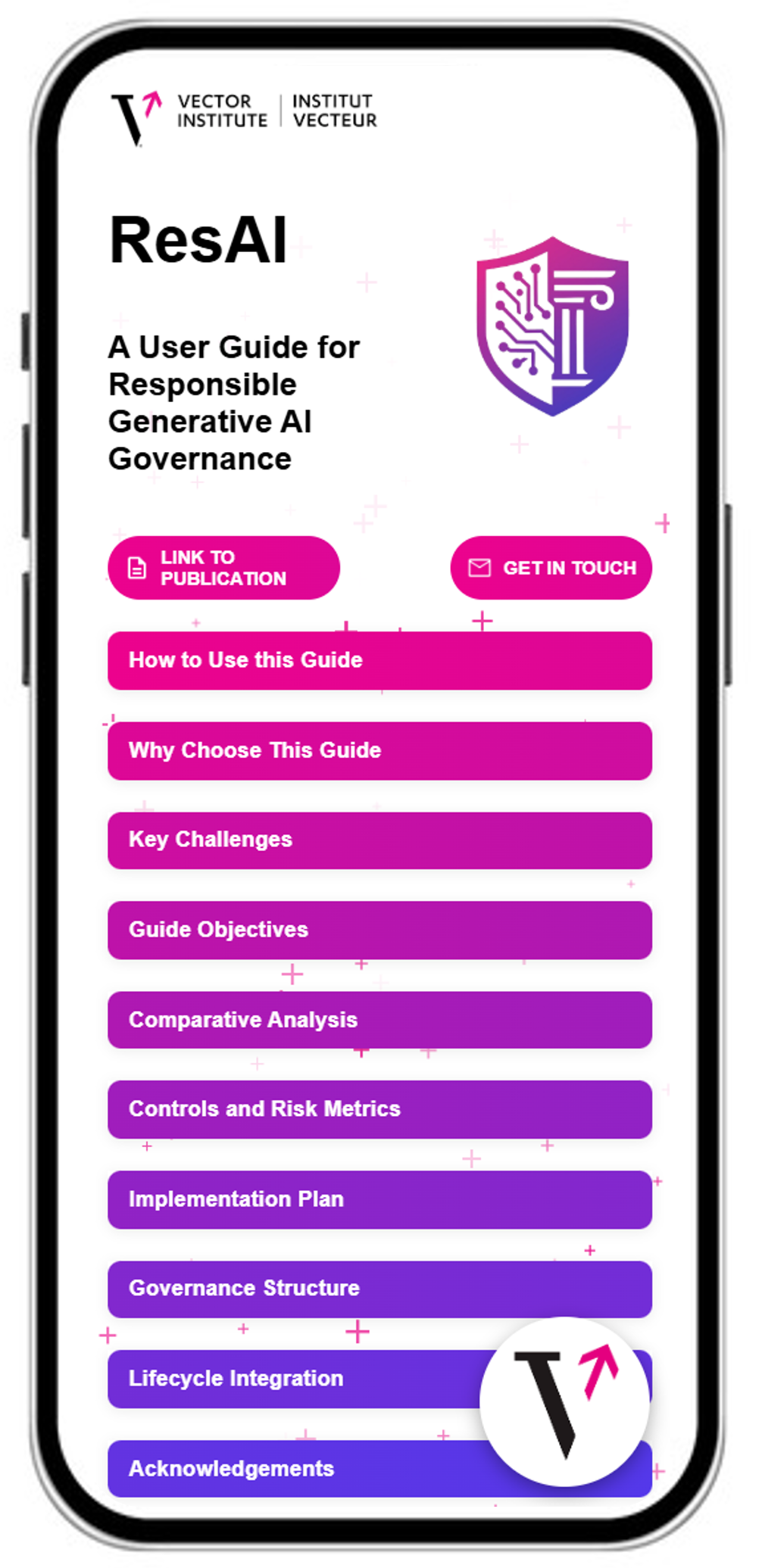}
    \caption{Responsible Generative AI (ResAI) Governance Website as an Online Resource https://res-ai.ca/~\cite{Hartman2025governanceagentdemo}}
    \label{fig:res-ai}
\end{figure}

As part of this work, we have developed an interactive Responsible GenAI Governance Guide (ResAI) that follows the structured steps outlined in Section~\ref{Implementation}. This guide is accessible as an online resource and is designed to support organizations in navigating responsible GenAI governance frameworks. Users can engage with the GenAI-powered chatbot, shown in Fig.~\ref{fig:res-ai}, which provides contextualized responses and guidance based on key governance principles and industry best practices. The ResAI Guide can be accessed at~\cite{Hartman2025governanceagent}, and the corresponding online repository is available at~\cite{Hartman2025governanceagentdemo}.

\section*{Acknowledgment}
This work is the result of a collaborative effort involving contributions from industry leaders, researchers, and practitioners across various domains. We extend our deepest gratitude to our co-authors and industry experts whose insights, expertise, and engagement have been instrumental in shaping this white paper.

We would like to express our gratitude to the following co-authors and industry leaders for their contributions to this work: Brian Ritchie, Fion Lee-Madan, Kathrin Gardhouse, Kathy Zielinski, Maggie Arai, Maitreya Kadam, Mark Paulsen, McKenzie Lloyd-Smith, Peter Slattery, Rob Straker, Susan Lindsay, Tiffany Wong, Andres Rojas, Sedef Kocak, Diana Moyana, Tahniat Khan, Veronica Chatrath, and Carolyn Chong. Their invaluable insights, thoughtful discussions, and expertise have enriched this work.

Furthermore, we acknowledge the institutions and documents referenced in this paper, whose research, frameworks, and guidance have informed our analysis and recommendations:

\begin{itemize}
    \item MIT Future Tech Lab - \url{https://airisk.mit.edu/}
    \item NIST AI Risk Management Framework - \url{https://www.nist.gov/itl/ai-risk-management-framework}
    \item Alan Turing Institute - \url{https://www.turing.ac.uk/research/research-projects/ai-ethics-and-governance-practice}
    \item UC Berkeley -
    \url{https://re-ai.berkeley.edu/sites/default/files/responsible_use_of_generative_ai_uc_berkeley_2025.pdf}
    \item Additional relevant documents as cited within this paper.
\end{itemize}

We acknowledge any copyright-related terms associated with the referenced materials. We appreciate the broader research community and governance professionals whose foundational work has informed our insights and recommendations. Their collective efforts continue to drive progress in AI governance, ensuring that technological advancements align with ethical, legal, and societal imperatives.

\bibliographystyle{IEEEtran} 
\bibliography{references}
\end{document}